\documentclass[12pt]{iopart}

\usepackage[dvipsnames]{xcolor}
\usepackage{tcolorbox}

\usepackage{iopams}  

\begin{document}

\title{Environmental thresholds for mass-extinction events }

\author{Guy R. McPherson$^1$, Beril Sirmacek$^2$*, Ricardo Vinuesa$^3$}
\address{$^1$School of Natural Resources and the Environment, University of Arizona, Tucson, Arizona, USA (e-mail: guy.r.mcpherson@gmail.com)\\
$^2$ Smart Cities, School of Creative Technology, Saxion University of Applied Sciences, The Netherlands (e-mail: b.sirmacek@saxion.nl)\\
$^3$ FLOW, Engineering Mechanics, KTH Royal Institute of Technology, Stockholm, Sweden and AI Sustainability Center, Stockholm, Sweden (e-mail: rvinuesa@mech.kth.se)}
\vspace{10pt}
\begin{indented}
\item[]September 2021
\end{indented}

\begin{abstract}
While the global-average temperatures are rapidly rising, more researchers have been shifting their focus towards the past mass-extinction events in order to show the relations between temperature increase and temperature thresholds which might trigger extinction of species. These temperature and mass-extinction relation graphs are found practical by conservationists and policy makers to determine temperature threshold values to set climate targets. Unfortunately, this approach might be dangerous, because mass-extinction events (MEEs) are related to many environmental parameters and temperature is only one of them. Herein we provide a more comprehensive evaluation of the environmental thresholds required to sustain a habitable planet. Besides, we suggest actions within the sustainable-development goals (SDGs) to observe those critical environmental parameters, in order to assure having an inhabitable planet for the current living species.
\end{abstract}

%
\vspace{2pc}

\section{Introduction}

Song et al.~\cite{song2021} published "Thresholds of temperature changes for mass extinctions" on 4 August 2021. They described the correlation between an increase in global average temperature and mass-extinction events (MEEs). The reference paper analyzed magnitudes and rates of temperature change and extinction rates of marine fossils during the past 450 million years. It discussed that MEEs during this time are highly correlated with global temperature changes, and it provided temperature thresholds which cause extinction of species. The reference paper concluded that a temperature increase of $5.2^{\circ}$C above the pre-industrial level at the present rate of increase would likely result in a MEE. We believe that there are more environmental parameters which are determining the MEEs and if accidentally this given temperature threshold value is set as a climate goal for our current situation, this would lead many living species - including humans - into catastrophic scenarios. Thus, herein we highlight that the global temperature $5.2^{\circ}$C alone cannot represent a trigger for a MEE. Much lower global temperature values are enough to trigger a MEE. Many other environmental parameters and the rate of environmental change are important factors limiting the adaptive response of organisms (e.g., Strona and Bradshaw \cite{strona2018}). We aim to draw attention to the importance of factors beyond a rise in temperature that pose an existential threat and we wish investigation of these factors to be taken into account within the Sustainable Development Goals (SDGs).

\section{Background}

MEEs do not result only from increases in global average temperature magnitude, as pointed out recently by McPherson \cite{McPherson2021} more environmental parameters and ecological dependencies must be considered. Not only is Earth currently in the midst of a MEE, but factors such as further increases in industrial activity, reductions in industrial activity (losing the aerosol masking), release of methane into the atmosphere, a decline in the harvest of grains, the looming ice-free Arctic Ocean, and vortices created by commercial aircraft can trigger abrupt loss of habitat for humans on Earth. Most scientists are unaware of the aerosol masking effect, which results from ongoing industrial activity. The aerosol masking effect, or global dimming, has been described in the peer-reviewed literature for at least 90 years \cite{Angstrom}. Coincident with industrial activity adding to greenhouse gases that warm the planet, industrial activity simultaneously cools the planet by adding aerosols to the atmosphere. Loss of aerosol masking will contribute to a rapid and substantial rise in global-average temperatures as indicated by Jia et al. \cite{Jia2021}

In addition to factors pointed out by McPherson \cite{McPherson2021}, volcanic eruption could cause rapid loss of habitat for our species. For example, the Toba super-volcano eruption caused severe tropical stratospheric ozone depletion \cite{Osipov2021}. Super-eruptions have been assumed to cause so-called, "volcanic winters" that act as primary evolutionary factors through ecosystem disruption and famine; however, winter conditions alone may not be sufficient to cause such disruption. Stratospheric sulfur emissions from the Toba super-eruption about 74,000 years ago caused severe stratospheric ozone loss through a radiation-attenuation mechanism. The Toba plume strongly inhibited oxygen photolysis, thereby suppressing ozone formation in the tropics, where exceptionally-depleted ozone conditions persisted for more than a year. This effect, when combined with the volcanic winter in the extra-tropics, could trigger a MEE \cite{Osipov2021}. Atri \cite{atri2019} showed that several exoplanets lie within the "habitable zones" of stars, where planets are thought to be capable of retaining liquid water on their surface, and therefore have the potential to host life. However, an exoplanet that is too close to its host star is highly sensitive to radiation bursts from the star, also known as flares. Atri found that "Solar proton events on close-in terrestrial planets can significantly enhance the radiation dose and adversely impact their habitability". Garcia-Sage et al. \cite{garcia2017} explained that Proxima b (an Earth-sized planet right outside our solar system in the habitable zone of its star) likely would be unable to retain its atmosphere in the wake of such an event, leaving the surface exposed to harmful stellar radiation and reducing its potential for habitability. Proxima Centauri’s powerful radiation drains the Earth-like atmosphere as much as 10,000 times faster than what occurs on Earth. As indicated by Strona and Bradshaw \cite{strona2018}, the ensuing rapid rate of environmental change likely to result from such an event would be sufficient to cause the loss of all life on Earth. McPherson \cite{McPherson2021} points out that the uncontrolled meltdown of nuclear facilities around the globe appears sufficient to produce the same dire outcome.

Naafs et al. \cite{Naafs2016} illustrates the rapid rise of contemporary changes in atmospheric carbon dioxide compared to past events, even those long-thought to be characterized by rapid change. “During the Aptian Oceanic Anoxic Event 1a, about 120 million years ago, … The rise of CO2 concentrations occurred over several tens to hundreds of thousand years.” Contrary to the notion that this event transpired very quickly, according to the lead author of the paper: “The change, however, appears to have been far slower than that of today, taking place over hundreds of thousands of years, rather than the centuries over which human activity is increasing atmospheric carbon dioxide levels.” In other words, “rapid” in the fossil record is nothing compared to today.

Hansen \cite{hansen2007} stated by analyzing the Plio-Pleistocene record; "With global warming of only $2-3 ^{\circ}$C and CO2 perhaps 350-450 ppm it was a dramatically different planet without Arctic sea ice in the warm seasons and with a sea level $25\mp10$ m higher."

Wasdell’s analysis from May 2014 \cite{Wasdell2014} explains the earth sensitivity with the following words; "the Global Climate System is extremely sensitive to small changes in average
surface temperature, much more so than had been understood when the policy target of not
exceeding a rise of $2^{\circ}$C was proposed and accepted by the international community. Dangerous
climate change is already with us as a consequence of a change of only $0.85^{\circ}$C. \textbf{Treating the
$2^{\circ}$C target as the boundary of “safe climate change” is an illusion which must now be
abandoned and replaced with a more realistic ceiling of not more than 1°C above the preindustrial level.}

Wasdell’s analysis from September 2015 \cite{Wasdell2015} includes several noteworthy conclusions: (1) “Current computer estimates of Climate Sensitivity are shown to be dangerously low,” revealing (2) “an eight-fold amplification of CO2 forcing (in contrast to the three-fold amplification predicted by the IPCC climate modelling computer ensemble), (3) \textbf{“the 2°C target temperature limit is set far too high”} (emphasis in original), and (4) “anthropogenic change is at least 100 times faster than at any time in the Paleo record.” The report’s bottom line: “There is no available carbon budget. It is already massively overspent, even for the 2°C target.” Glikson \cite{Glikson2020} has also confirmed with these words "During the Anthropocene greenhouse gas forcing has risen by more than $2.0 W/m^2$, which constitutes an abrupt event of a period not much longer than a lifetime."

\section{Discussion}

In the light of some of the key literature elements that we have presented, the following subsections include our major comments as a response to the reference article titled "Thresholds of temperature change for mass extinctions" \cite{song2021}.


{\bf Comment 1: The article provides an incomplete analysis of extinction rates based purely on a rise in global temperature.}

The authors showed their results as if the mass-extinction rates are simply correlated with changes in temperature. However, the parameters listed below would also contribute to a MEE. Some of these parameters may have contributed to previous MEEs. It is not known whether some of these other parameters were primary contributors to prior MEEs.

\begin{tcolorbox}

\textbf{Other major parameters which would create existential threat (some of which might have contributed to previous MEEs):} Change of the CO$_2$ and methane levels in the atmosphere, change of the ultraviolet (UV) stress, change of the water and soil acidification, ozone-layer changes, change of air humidity, change of basic-life resources, nuclear activities.

\end{tcolorbox}









{\bf Comment 2: The article compares the extinction rates simply based on temperature changes, thereby ignoring the rapidity of these changes.}

Slow temperature changes will provide opportunities for species to adapt. However, the rapidity of environmental change produced by abrupt climate change is fundamentally more important than the magnitude of the change alone. Therefore, not only the magnitude changes, but the first- and the second-order derivatives of those changes (speed and acceleration) are critical to understanding extinction rates.


{\bf Comment 3: Land warms faster than water.}

Terrestrial and aquatic bodies warm and cool at different rates. A global temperature rise above $2^\circ$C might cause mass-extinction of land animals (including humans). However, the same magnitude of warming might not produce such an outcome for sea animals. Thus, a comparison of extinction rates comparison must be applied separately to terrestrial and aquatic water and land animals.


{\bf Comment 4: The article compares extinction rates from different time periods regardless of the climate-adaptation capabilities of the species within each extinction event.}

Different species are driven to extinction via different means. For example, marine animals die from acidification or decreased oxygen even even when they have sufficient food. On the other hand, terrestrial animals might lose habitat when an abrupt heating event destroys sources of food. Thus, the causes of extinction within various groups must be classified before comparisons are made. Different animals also possess different ecological dependencies: the extinction of one species might directly result in the extinction of another species, even when the latter would have been able to adapt to the temperature or other changes, a phenomenon labeled, "co-extinctions" by Strona and Bradshaw  \cite{strona2018}.

\section{Conclusions} 


Based on the comments listed above, we believe that the article ``Thresholds of temperature change for mass extinctions'' \cite{song2021} fails to provide a comprehensive understanding of MEEs. At the very least, comparisons of different MEEs can be expanded via consideration of additional contributing factors beyond planetary heating.


The analysis by Song et al. indicates a stunningly high temperature threshold for humans on Earth. This high global-average temperature increase assumes temperatures below this threshold are safe for humans, which seems unlikely in light of the ongoing and projected rates of environmental change. Importantly, the framework of the Sustainable Development Goals (SDGs)~\cite{UNGA}, requires a holistic view to understand the interconnections among goals and targets~\cite{VinuesaNat}. The Sendai Framework for Disaster-Risk Reduction~\cite{sendai}, which is extremely relevant for SDG 13 (on climate change), clearly states the need for multi-hazard and multi-sector disaster-risk reduction practices, together with close interactions between private and public actors and an emphasis on research institutions. The need for regulation of aspects beyond temperature levels is also reflected in SDG 14 (on life below water), for instance in the context of reducing ocean acidification, and in SDG 15 (on life on land), where forest management and biodiversity are among the indicators of SDG achievement. Importantly, these three SDGs include the adoption of specific regulations (global and local) targeting a wide range of indicators beyond global temperature.

Therefore, we urge scientists to reconsider safe conditions for habitability of Earth for our species and the many others on which we depend for our survival. Such an approach must be based on a comprehensive assessment of environmental parameters well beyond changes in global warming.

Besides, since Earth has already lost a massive amount of the cryosphere important to maintaining albedo, gradual efforts to sequester atmospheric carbon and reduce greenhouse gas emissions appear inadequate to retain habitat for our species. Given our current environmental situation, some type of geoengineering project appears necessary if we are to retain habitat for our species.

\section*{ACKNOWLEDGEMENTS}\label{ACKNOWLEDGEMENTS}

Author contributions are represented in the author order. RV acknowledges financial support from the Swedish Research Council (VR).

\section*{References}

\bibliography{journal}

\end{document}